\documentstyle[sprocl,epsfig]{article}

\def\be{\begin{equation}}
\def\ee{\end{equation}}
\def\bea{\begin{eqnarray}}
\def\eea{\end{eqnarray}}

\def\sqee{\sqrt{s}_{\rm ee}}
\def\sqeeb{\sqrt{s}_{\protect\bf\rm ee}}

\def\ee{\mbox{e}^+\mbox{e}^-}

\def\WECAL{W_{\rm ECAL}}

\newcommand{\sleq} {\raisebox{-.6ex}{${\textstyle\stackrel{<}{\sim}}$}}

\def\ETJET{E^{\rm jet}_T}

\def\ETMIN{E^{\rm min}_T}

\def\xgp{x_{\gamma}^+}
\def\xgm{x_{\gamma}^-}
\def\xgpm{x_{\gamma}^{\pm}}
\def\etajet{\eta^{\rm jet}}

\def\Zzero{\ifmmode {{\mathrm Z}^0} \else {${\mathrm Z}^0$} \fi}

\def\cost{\cos\theta^*}

\begin{document}

\title{DIJET PRODUCTION IN PHOTON-PHOTON COLLISIONS \\
 AT \boldmath $\sqeeb=161$ AND \boldmath $172$~GEV \footnotemark[1]}

\author{ ROLAND B\"URGIN \\ for the OPAL collaboration}

\address{Universit\"at Freiburg, Hermann-Herder-Stra\ss e 3,\\
D-79104 Freiburg, Germany }


\maketitle\abstracts{Dijet production is studied in collisions
of quasi-real photons radiated by the LEP beams at e$^+$e$^-$
centre-of-mass energies $\sqee=161$ and $172$ GeV. Jets are
reconstructed using a cone jet finding algorithm in the range
$|\etajet|<2$ and $\ETJET>3$~GeV. 
The angular distributions of direct and double-resolved processes
and the inclusive two-jet cross-section are measured. They are compared
to next-to-leading order perturbative QCD calculations and the
prediction of the leading order Monte Carlo generators PYTHIA
and PHOJET.}

\section{Introduction}
\footnotetext[1]{To be published in the proceedings of PHOTON'97, 
Egmond aan Zee}
The production of dijet events in the collision of two quasi-real photons
is used to study the structure of the photon and different QCD predictions.
The measurement of inclusive jet cross-sections and the comparison
with next-to-leading order (NLO) QCD calculations
and different photon structure functions can constrain the relatively
unknown gluonic content of the photon.

\vspace{-1mm}
\section{Event selection and jet finding}
\label{sec-evsel}

To select a data sample of two-photon events the following set of cuts
was applied.
The sum of all energy deposits in the electromagnetic calorimeter
(ECAL) and the hadronic calorimeter (HCAL) has to be less than 45 GeV.
The visible invariant hadronic mass, $\WECAL$, measured
in the ECAL has to be greater than 3 GeV.
The missing transverse energy of the event
measured in the ECAL and the forward
calorimeters has to be less than 5 GeV.
At least 5 tracks must have 
been found in the tracking chambers.
Events with detected scattered electrons (single-tagged or
double-tagged events) are excluded from the analysis.

The results of the cone jet finding algorithm depend on the minimal
transverse energy $\ETMIN$ and the cone size
$R=\sqrt{(\Delta\eta)^2+(\Delta\phi)^2}$ with the
pseudorapidity $\eta=-\ln\tan(\theta/2)$ and the azimuthal
angle $\phi$.
Here the values were chosen to be $R=1$ and $\ETMIN=3$~GeV.
The jet pseudorapidity in the laboratory frame is required to
be within $|\etajet|<2$.

After applying all cuts and requiring at least
two jets 2681 events remain in the data corresponding to an integrated
luminosity of 20 pb$^{-1}$.
For the simulation of two-photon interactions the Monte Carlo generators
PYTHIA \cite{bib-pythia} and PHOJET \cite{bib-phojet} have been used.
The mean $Q^2$ of the selected Monte Carlo events is 0.06~GeV$^2$.
About 1.2~\% of the events in the data sample are expected to be
$\ee$ annihilation events with hadronic final states and 0.2~\%
electron-photon events.

\section{Angular distributions in direct and resolved events}
\label{sec-angdis}

A pair of variables, $\xgp$ and $\xgm$, can be defined
which is related to the fraction of the photon energy participating in the
hard scattering:
\begin{equation}
\xgp=\frac{\displaystyle{\sum_{\rm jets=1,2}(E+p_z)}}
 {{\displaystyle\sum_{\rm hadrons}(E+p_z)}} \;\;\;\mbox{and}\;\;\;
\xgm=\frac{\displaystyle{\sum_{\rm jets=1,2}(E-p_z)}}
{\displaystyle{\sum_{\rm hadrons}(E-p_z)}},
\label{eq-xgpm}
\end{equation}
where $p_z$ is the momentum component along the $z$ axis of the
detector and $E$ is the energy of the jets or hadrons.
If an event contains more than two jets, the two jets with
the highest $\ETJET$ values are taken. 
Samples with large direct and double-resolved contributions
can be separated by requiring both $\xgp$ and $\xgm$ to be
larger than 0.8 (denoted as $\xgpm > 0.8$) or both
values to be smaller than 0.8 (denoted as $\xgpm < 0.8$),
respectively \cite{bib-opalgg}.

\begin{figure}[htbp]
   \begin{center}
      \mbox{
          \epsfxsize=11.cm
          \epsffile{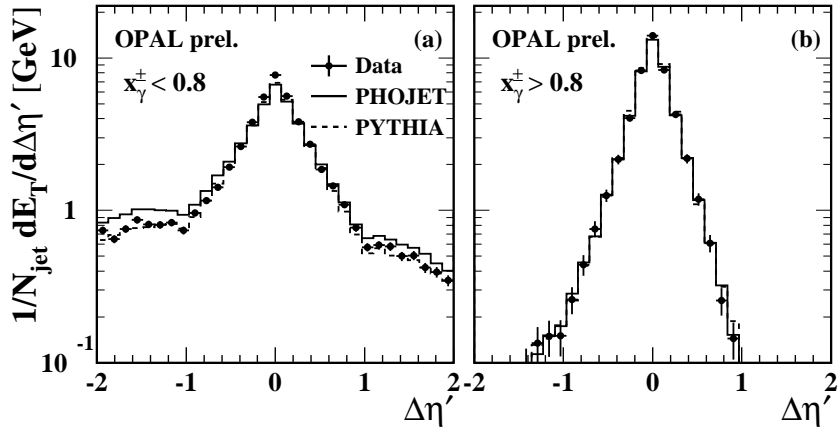}
           }
   \end{center}
\caption{Uncorrected energy flow transverse to the beam direction
measured relative to the direction of each jet in two-jet events. 
Jets from data samples with a large contribution of (a) double-resolved
and (b) direct events according to their $\xgp$ and $\xgm$ values are
shown. The energy flow is integrated over $|\Delta\phi|<\pi/2$.
Statistical errors only are shown.}
\label{fig-jetprof}
\end{figure}

The transverse energy flow around the jet axis for double-resolved
events is expected to show additional activity outside the jet
due to the photon remnants compared to the direct events.
Figure~\ref{fig-jetprof} therefore shows the transverse energy flow
around the jets with respect to the jet direction for data samples
with different $\xgpm$ cuts. The pseudorapidity difference
is defined by:
$$\Delta\eta'=k(\eta-\etajet).$$
The factor $k$ is chosen event-by-event to be
$k=+1$ for events with $\xgp>\xgm$ and $k=-1$ for
events with $\xgp<\xgm$.
Due to the definition of $\Delta\eta'$
there is always more of the remnant found at $\Delta\eta'<0$
and the enhancement due to the additional transverse energy flow 
at negative and positive $\Delta\eta'$ is asymmetric. 
The jets in the data sample with $\xgpm > 0.8$ are more collimated
and there is almost no activity outside the jet whereas the
transverse energy flow of two-jet events with $\xgpm < 0.8$
shows additional activity outside the jets.

In the centre-of-mass system of the interacting partons or bare photons
the parton scattering angle $\theta^*$ is defined as the angle between
the jet axis of the jets originating from the outgoing partons
and the axis of the incoming partons or bare photons.
The angular distribution of the jets can be used 
to check the separation between the different event classes using
a variable which is not directly correlated to the definition of $\xgp$
and $\xgm$.
In the dijet centre-of-mass frame one expects different angular
distributions for direct and double-resolved events.
An estimator of the angle between the jets and the parton-parton
axis in the dijet centre-of-mass frame can be formed from their
pseudorapidities. 
The variable $\cost$ is calculated as
$$\cost=\tanh\left(\frac{\eta^{\rm jet1}-\eta^{\rm jet2}}{2}\right).$$
Since the ordering of the jets is arbitrary, only $|\cost|$ can be measured. 
The matrix elements of elastic parton-parton scattering processes
have been calculated in LO \cite{bib-maelm}.
The cut on $\ETJET$ restricts the accessible range of values
of $|\cost|$. 
Requiring the invariant mass of the dijet system to be larger than
12~GeV ensures that values of $|\cost|<0.85$ are not biased
by the $\ETJET$ cut.
The boost of the two-jet system in the $z$ direction is defined by
$\bar{\eta}=(\eta^{\rm jet1}+\eta^{\rm jet2})/2.$
The detector resolution on $|\cost|$ deteriorates significantly for
events with $|\bar{\eta}|$ larger than 1.
These events were therefore rejected by requiring $|\bar{\eta}| < 1$.
\begin{figure}[htbp]
   \begin{center}
      \mbox{
          \epsfxsize=11.0cm
          \epsffile{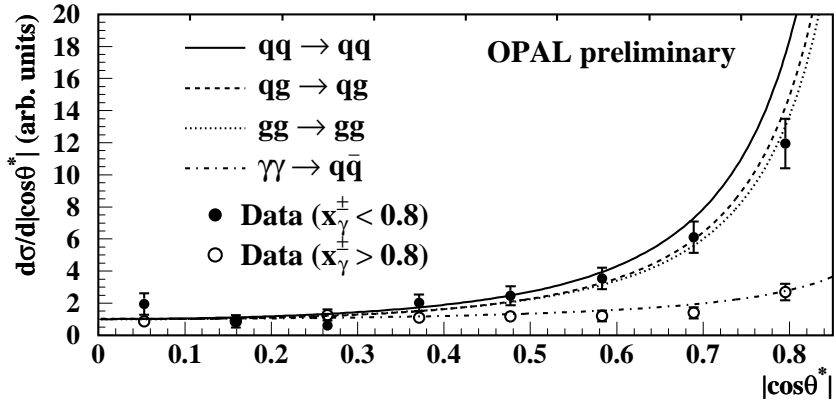}
           }
   \end{center}
\caption{Angular distribution of events with large direct and
large double-resolved contributions according to the separation
with $\xgp$ and $\xgm$. The data are compared to QCD matrix element
calculations \protect\cite{bib-maelm}. 
The data are normalised
to have an average value of 1 in the first three bins and the curves
are normalised to be 1 at $\cos(0)$.}
\label{fig-costhst}
\end{figure}

Figure \ref{fig-costhst} shows the
$\left|\cost\right|$-distribution of events with $\xgpm>0.8$ and
of events with $\xgpm < 0.8$. 
The dependence on the Monte Carlo models used is taken
into account by adding the difference between the results obtained
with PYTHIA, which are taken to be the central values, and PHOJET
to the systematic error.
The  error bars show the statistical and the systematic errors
added in quadrature.
The events with
$\xgpm > 0.8$ show a small rise with $|\cost|$, whereas the
events with $\xgpm < 0.8$ show a much stronger rise in $|\cost|$.
The data points of the events with $\xgpm<0.8$ are compared
with the prediction of a QCD matrix element calculation 
of the interaction of quarks or gluons in the photon.
The matrix elements with a relevant contribution to the
cross-section where antiquarks are involved instead of quarks
show a similar behaviour as the examples shown.
The QCD matrix element
calculations agree well with the data points of the data samples with
large direct and large double-resolved contribution.

\section{Inclusive two-jet cross-sections}
\label{sec-cross}
\begin{figure}[htbp]
   \begin{center}
      \mbox{
          \epsfxsize=11.0cm
          \epsffile{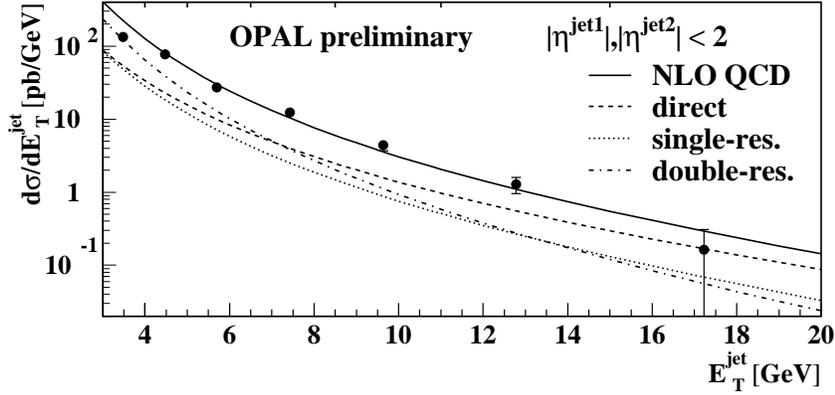}
           }
   \end{center}
\caption{The inclusive two-jet cross-section as a function
of $\ETJET$ for jets with $|\etajet|<2$ compared to the NLO
calculation by Kleinwort and Kramer \protect\cite{bib-kleinwort}.
The direct, single-resolved and double-resolved cross-sections
and the sum (continuous line) are shown separately.}
\label{fig-ettwojet}
\end{figure}

The inclusive two-jet cross-section is measured using a
cone jet finding algorithm with a cone size $R=1$.
The data were corrected for the selection cuts, the resolution effects
of the detector and the background from non-signal processes. 
In Fig.~\ref{fig-ettwojet}, the inclusive two-jet cross-section is
shown as a function of $\ETJET$.
The bin sizes, which are indicated by the vertical lines at the top
of the figure, approximately reflect the experimental resolution.

The $\ETJET$ distribution is compared to an NLO
perturbative QCD calculation of the inclusive two-jet cross-section
by Kleinwort and Kramer \cite{bib-kleinwort} who use
the NLO GRV parametrisation of the photon structure function~\cite{bib-grv}.
Their calculation was repeated for the kinematic conditions
of this analysis.
The direct, single- and double-resolved parts and their sum are
shown separately. The resolved
cross-sections dominate in the region $\ETJET\;\sleq\;8$~GeV,
whereas, at high $\ETJET$ the direct cross-section is largest.
The data points are in good agreement with
the calculation except in the first bin where the NLO calculation is
not reliable due to IR singularities.
The uncertainties due to the modelling of the hadronisation process
are expected to contribute mainly at low $\ETJET$ values.
\begin{figure}[htbp]
   \begin{center}
      \mbox{
          \epsfxsize=11.0cm
          \epsffile{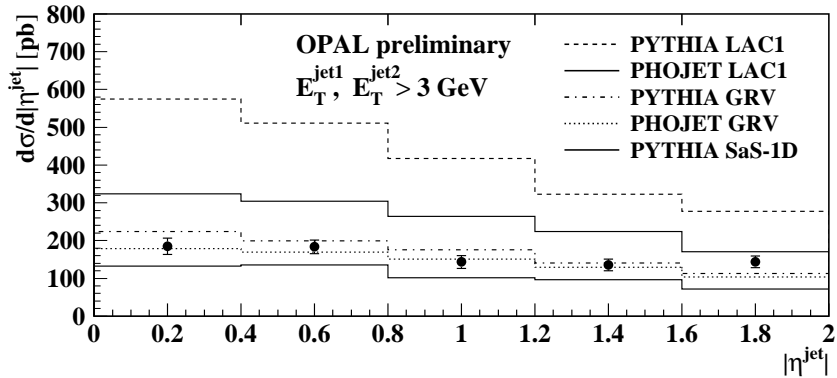}
           }
   \end{center}
\caption{The inclusive two-jet cross-section as a function of $|\etajet|$
for jets with $\ETJET > 3$~GeV.}
\label{fig-etatwojet}
\end{figure}

The inclusive two-jet cross-section, which is dominated by the low
$\ETJET$ events, depends on the parton density function of the photon
which mainly differs in the assumptions on
the gluon part of the photon. In 40 to 60 \% of all processes at
least one gluon from the photon interacts depending on the
parton density function used. This leads to different predictions
of the inclusive two-jet cross-section.
The inclusive two-jet cross-section as a function of $|\etajet|$
is shown in Fig.~\ref{fig-etatwojet}.
The inclusive two-jet cross-section predicted by the two Monte Carlo
models differ significantly even if the same photon structure function
is used. This model dependence currently reduces the
sensitivity to the parametrisation of the photon structure function.
The GRV-LO and SaS-1D parametrisations~\cite{bib-grv,bib-sas} describe
the data equally well, but the LAC1 parametrisation \cite{bib-LAC1}
overestimates the inclusive two-jet cross-section
significantly.

\section{Conclusions}
\label{sec-conclusions}
The distribution of the parton scattering angle $\theta^*$ 
of data samples with large direct and double-resolved contributions 
separated experimentally using the variables $\xgp$ and $\xgm$
have been compared to the relevant QCD matrix element calculations. 
The inclusive two-jet cross-sections were measured as a function
of $\ETJET$ and $|\etajet|$.
The $\ETJET$ dependent two-jet cross-section is in good agreement
with an NLO QCD calculation.
The GRV-LO and SaS-1D parametrisations describe the inclusive two-jet
cross-section equally well. The LAC1 parametrisation
overestimates the inclusive two-jet cross-section
significantly.

\end{document}